\documentclass[a4paper]{article}

\usepackage[pages=all, color=black, position={current page.south}, placement=bottom, scale=1, opacity=1, vshift=5mm]{background}
\SetBgContents{
}      

\usepackage[margin=1in]{geometry} 

\usepackage{amsmath}
\usepackage{amsthm}
\usepackage{amssymb}

\usepackage[utf8]{inputenc}
\usepackage{hyperref}

\usepackage{cite}
\usepackage{amsmath,amssymb,amsfonts}
\usepackage{algorithmic}
\usepackage{graphicx}
\usepackage{textcomp}
\usepackage{xcolor}
\usepackage{multirow}
\usepackage{subcaption}
\usepackage{braket}
\usepackage{comment}
\usepackage{enumitem}
\hypersetup{
	unicode,
	pdfauthor={Author One, Author Two, Author Three},
	pdftitle={A simple article template},
	pdfsubject={A simple article template},
	pdfkeywords={article, template, simple},
	pdfproducer={LaTeX},
	pdfcreator={pdflatex}
}

\usepackage[sort&compress,numbers,square]{natbib}
\theoremstyle{plain}

\theoremstyle{definition}

\graphicspath{{fig/}}

\usepackage{mathrsfs} 

\title{Quantum Circuit Simulation by SGEMM Emulation on Tensor Cores and Automatic Precision Selection}
\author{Hiroyuki Ootomo$^1$\and
Hidetaka Manabe$^2$ \and
Kenji Harada$^2$ \and
Rio Yokota$^1$
}

\date{
	$^1$Tokyo Institute of Technology \\ \texttt{ootomo.h@rio.gsic.titech.ac.jp, rioyokota@gsic.titech.ac.jp}\\%
	$^2$Kyoto University \\ \texttt{manabe@acs.i.kyoto-u.ac.jp, harada.kenji.8e@kyoto-u.ac.jp}\\[2ex]%
}

\begin{document}
 \maketitle

 \begin{abstract}
 Quantum circuit simulation provides the foundation for the development of quantum algorithms and the verification of quantum supremacy.
Among the various methods for quantum circuit simulation, tensor network contraction has been increasing in popularity due to its ability to simulate a larger number of qubits. 
During tensor contraction, the input tensors are reshaped to matrices and computed by a GEMM operation, where these GEMM operations could reach up to 90\% of the total calculation time.
GEMM throughput can be improved by utilizing mixed-precision hardware such as Tensor Cores,
but straightforward implementation results in insufficient fidelity for deep and large quantum circuits.
Prior work has demonstrated that compensated summation with special care of the rounding mode can fully recover the FP32 precision of SGEMM even when using TF32 or FP16 Tensor Cores.
The exponent range is a critical issue when applying such techniques to quantum circuit simulation.
While TF32 supports almost the same exponent range as FP32, FP16 supports a much smaller exponent range.
In this work, we use the exponent range statistics of input tensor elements to select which Tensor Cores we use for the GEMM.
We evaluate our method on Random Circuit Sampling (RCS), including Sycamore's quantum circuit, and show that the throughput is 1.86 times higher at maximum while maintaining accuracy.

 \end{abstract}

 \section{Introduction}
Quantum circuit simulators are vital for the development of quantum algorithms and verification of \textit{quantum supremacy}, and are considered as one of the key applications for HPC systems in the Exa-scale era \cite{villalonga_establishing_2020,liu_closing_2021,nguyen_tensor_2021}.
In a quantum computer, all operations follow quantum mechanics: preparing qubits (a quantum version of classical bits), applying unitary gates, and measuring qubits to get classical data.
The goal of quantum circuit simulation is to reproduce the classical result obtained by these quantum operations only with a classical computer.

\begin{figure}[t]
    \centering
    \includegraphics[width=0.8\linewidth]{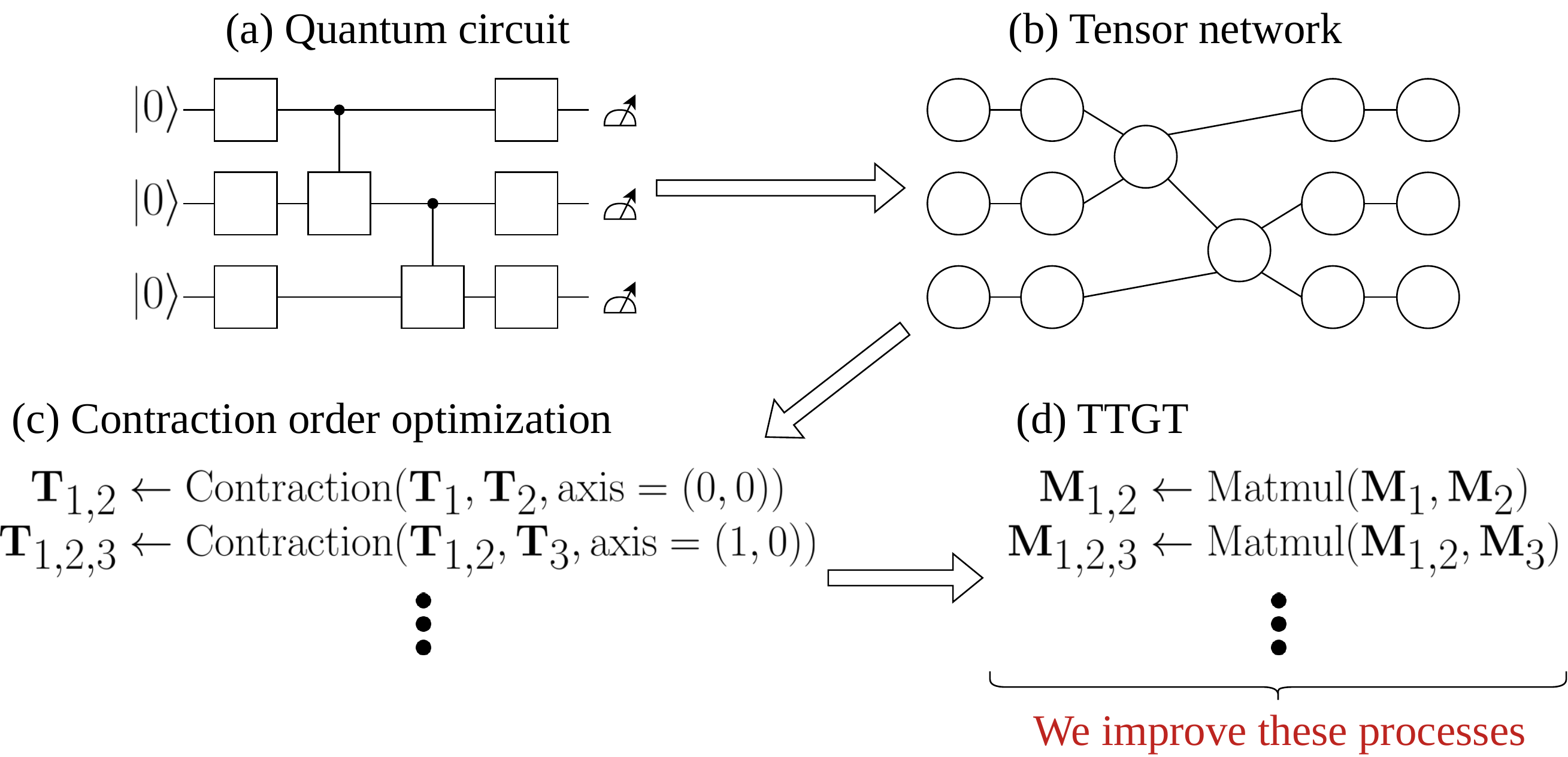}
    \caption{An image of quantum circuit simulation using tensor network simulation and the part of the computation that we improve.}
    \label{fig:qc-intro}
\end{figure}
There exist various types of quantum simulators\cite{suzuki_qulacs_2021,jones_quest_2019,treinish_qiskitqiskit_2022-1,guerreschi_intel_2020}. For general circuits dominated by non-Clifford gates, the two types of simulators: 1) state vector and 2) tensor network methods are widely used.
We choose these simulation methods according to the objectives.
The state vector simulations require $2^{n}$ complex values on memory, where $n$ is the number of qubits.
For instance, to simulate Google's Sycamore \cite{arute_quantum_2019-2}, which has 53 qubits, we would require 128 PB of memory.
Since the total memory capacity of the current largest supercomputers is in the order of a few PB, state vector methods are limited by memory capacity.
One advantage of state vector methods is that the computational complexity for a circuit of depth $d$ is $\mathcal{O}(2^{n} \times d)$ and scales linearly with $d$.
Furthermore, there are some studies to reduce the required memory size, such as by splitting the circuit \cite{chen_64-qubit_2018}.
On the other hand, the tensor contraction method \cite{markov_simulating_2008} can simulate several thousands of qubits with low-depth layers at a slightly higher computational cost.
Therefore, tensor contraction is the method of choice for many recent studies that aim to validate quantum supremacy in both quantum computing and high performance computing\cite{markov_quantum_2018,villalonga_establishing_2020,pan2022Simulation}.
In tensor contraction methods, the quantum circuit is represented as a tensor network, where each node represents a quantum gate, and the edge represents the quantum wire, as shown in Fig. \ref{fig:qc-intro}.
The appearance probability of an output bitstring is calculated through the contraction of the tensor network.
Since the computational complexity of the contraction heavily depends on its order of computation, many studies focus on finding the near-optimal contraction order \cite{huang2021Efficient,liang_fast_2021,gray_hyper-optimized_2021,pan2022Simulation}.

The simplest way to compute a tensor contraction is to form nested loops for each tensor index. In practice, the TTGT (Transpose-Transpose-GEMM-Transpose) algorithm is widely used since it can leverage the existing high performance GEMM implementations on various processors such as Intel MKL and NVIDIA cuBLAS.
In this algorithm, the input tensors are first reshaped into matrices, where rows or columns separate the contracted and non-contracted indices.
Then, it computes the matrix multiplication of the input matrices and transposes them to fit into the output tensor if necessary.
Although this approach requires additional memory to store the transposed tensors, it can leverage the high performance GEMM implementation, which can effectively utilize the hierarchical cache and memory of the target processor.

\begin{figure}[t]
    \centering
    \includegraphics[width=0.9\linewidth]{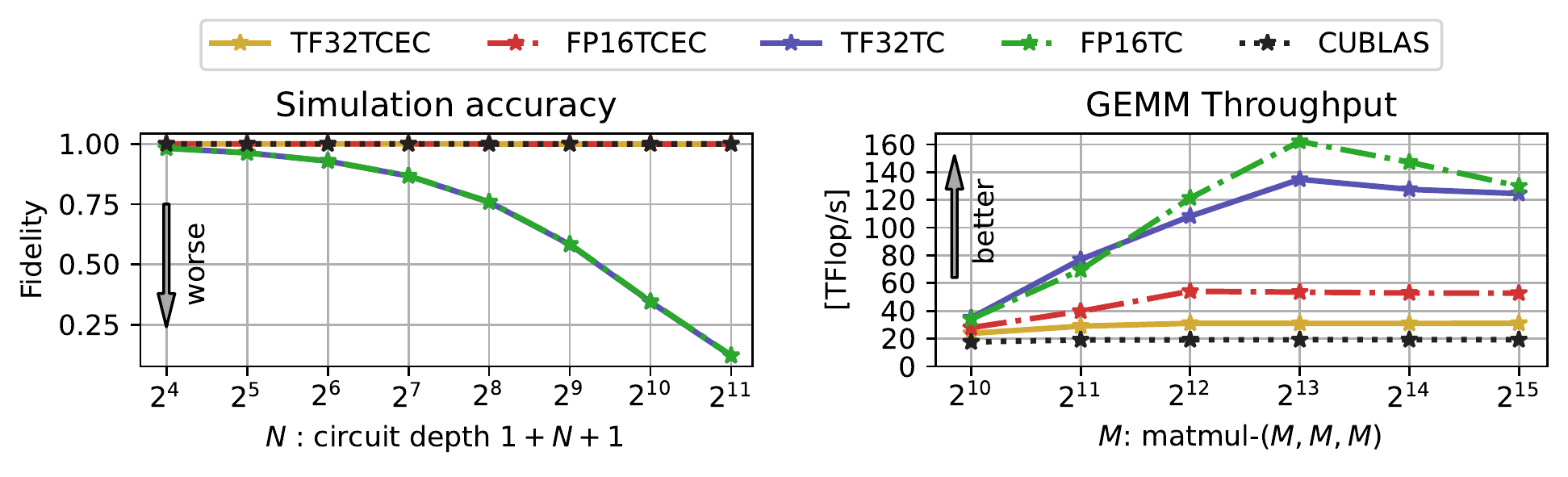}
    \caption{{\bf Left:} The comparison of simulation accuracy of a $4 \times 4$ rectangular lattice Random Quantum Circuit simulation for each GEMM precision. TF32TCEC, FP16TCEC, and CUBLAS have single-precision mantissa accuracy, and TF32TC and FP16TC have half-precision. The circuit consists of a Hadamard gate layer ($1+$), $N$-CZ gate layers, and a Hadamard gate layer $(+1)$. {\bf Right:} The throughput comparison of each CGEMM implementation.}
    \label{fig:intro-top}
\end{figure}
The chain of GEMM operations in tensor contraction can be accelerated by improving both the algorithm and implementation.
In terms of algorithmic improvement, Huang \textit{et al.}\cite{huang_implementing_2018} applied Strassen's algorithm to tensor contraction and slightly reduced its computational complexity.
With respect to implementation, leveraging the existing high performance GEMM implementations on Tensor Cores would seem like a natural fit, but quantum circuit simulations require at least single precision accuracy, as we will show later.
Tensor Cores are mixed-precision matrix multiply-add units on NVIDIA GPUs, and have $7.5\sim 15\times$ throughput compared to the standard arithmetic units on NVIDIA A100 GPUs.
Although the data type of input matrices for multiplication are low-precision (FP16 or TF32), the computation inside Tensor Cores is performed in higher-precision (FP32).
However, when we compute an SGEMM on Tensor Cores, we need to convert the input matrices to low precision, which causes accuracy degradation.
Markidis \textit{et al.} propose a method to recover the accuracy through compensated summation, but their method cannot fully recover the FP32 accuracy \cite{markidis_nvidia_2018}.
Our previous study identified the cause of this problem as the rounding mode inside the Tensor Cores, and developed a method that circumvents this issue with minimal performance degradation \cite{ootomo_recovering_2022}.
As a result, this method outperforms the theoretical peak performance of FP32 SIMT Cores on NVIDIA A100 GPUs while the FP32 accuracy is fully recovered.
This method can be applied to TF32 Tensor Core error correction (TF32TCEC) and FP16 Tensor Core error correction (FP16TCEC).
The TF32 version supports almost the same exponent range as FP32, while the FP16 version supports only a limited exponent range but has higher throughput.
Therefore, there is a trade-off between the supported exponent range and throughput.
In the case of quantum circuit simulation using tensor network contraction, it is difficult to compute the simulation in high precision on Tensor Cores without the error correction when the number of computations is large, as shown in Fig. \ref{fig:intro-top}.
Therefore, error correction is necessary when using Tensor Cores for the simulation.
However, it is difficult to determine which tensor contraction requires TF32TCEC or if FP16TCEC is sufficient.
To the extent of the authors' knowledge, there is no framework for automatically selecting between these two operations.
Although Liu \textit{et al.} use a dynamic scaling method for FP16 computation in the simulation, they select the parts to be computed with FP16 heuristically \cite{liu_closing_2021} by performing an analysis of the tensor network contraction a priori.
Furthermore, their method causes overflow in some cases, which causes the entire computation to fail.

In the present work, we drastically improve the throughput of quantum circuit simulation while retaining sufficient accuracy by using the SGEMM emulation on Tensor Cores and automatic precision selection.
To select between TF32TCEC and FP16TCEC for each tensor contraction, we use the exponent statistics of elements in the input matrices measured before the GEMM operation.

The summary of our contributions is as follows:
\begin{enumerate}
    \item We develop a library for SGEMM emulation on Tensor Cores, cuMpSGEMM, that can be used without any change to the source code of the target applications.
    This library intercepts SGEMM function calls of the cuBLAS dynamic library and executes the SGEMM emulation on Tensor Cores instead, and surpasses the performance of SGEMM while fully retaining the accuracy.
    This library is not limited to quantum circuit simulation and can be used for any other application that calls cuBLAS SGEMM, CGEMM, and their batched variants and is open-source and available on GitHub\footnote{\url{https://github.com/enp1s0/cuMpSGEMM}}.

    \item We develop a method to select the GEMM precision, TF32TCEC, FP16TCEC, or FP16TCEC with scaling, for improving the throughput by taking the exponent statistics of input matrix elements before the GEMM operation.
    We have tested our method on a random tensor network contraction and confirmed that it successfully avoids the underflow error with a slight overhead.
    \item We evaluate the accuracy and throughput of Random Circuit Sampling (RCS) simulation, including Sycamore's quantum circuit.
    In RCS, our method improves the throughput by $1.86$ times for a quantum circuit of $9 \times 9$, depth=$33$, and $1.44$ times for the Sycamore circuit while retaining the accuracy of the baseline implementation.
\end{enumerate}

\section{Background}

\subsection{NVIDIA Tensor Core and SGEMM emulation}
NVIDIA Tensor Core computes a matrix multiplication and addition,
\begin{equation}
    \mathbf{D}_\text{F32} \leftarrow \mathbf{A}_\text{low}\cdot\mathbf{B}_\text{low}+\mathbf{C}_\text{F32},
\end{equation}
where the subscript denotes the data type of the matrix: the ``low" is low-precision, FP16 or TF32, and ``F32" is FP32.
Although $\mathbf{A}_\text{low}$ and $\mathbf{B}_\text{low}$ are low-precision, the multiplications and additions are computed in FP32.
However, when it comes to computing single-precision matrix multiplication on Tensor Cores, we must convert the input matrices from single-precision to low-precision, which causes a loss of accuracy in the final computation result.
To recover the loss of accuracy, Markidis \textit{et al.} propose an error correction method based on compensated summation \cite{markidis_nvidia_2018}, but their method does not recover the full FP32 accuracy.
Our previous study improves upon this method by avoiding the rounding inside Tensor Cores and can recover the full FP32 accuracy with minimum overhead\cite{ootomo_recovering_2022}.
In our method, a single-precision matrix-multiplication $\mathbf{C}_\text{F32}\leftarrow\mathbf{A}_\text{F32}\cdot\mathbf{B}_\text{F32}$ is computed approximately as follows.
\begin{align}
    \mathbf{A}_\text{low} &\leftarrow \text{toLow}\left(\mathbf{A}_\text{F32}\right) \nonumber \\
    \Delta\mathbf{A}_\text{low} &\leftarrow \text{toLow}\left(\left(\mathbf{A}_\text{F32}-\text{toF32}\left(\mathbf{A}_\text{low}\right)\right)\times 2^{11}\right) \nonumber \\
    \mathbf{B}_\text{low} &\leftarrow \text{toLow}\left(\mathbf{B}_\text{F32}\right) \nonumber \\
    \Delta\mathbf{B}_\text{low} &\leftarrow \text{toLow}\left(\left(\mathbf{B}_\text{F32}-\text{toF32}\left(\mathbf{B}_\text{low}\right)\right)\times 2^{11}\right) \nonumber \\
    \label{eq:tcec-last}
    \mathbf{C}_\text{F32} &\approx \mathbf{A}_\text{low} \cdot \mathbf{B}_\text{low} + \left(\Delta\mathbf{A}_\text{low} \cdot \mathbf{B}_\text{low} + \mathbf{A}_\text{low} \cdot \Delta\mathbf{B}_\text{low}\right)/2^{11},
\end{align}
where ``toLow" is the conversion from FP32 to low-precision and ``toF32" is from low-precision to FP32.
In this scheme, each matrix element in $\mathbf{A}_\text{F32}$ and $\mathbf{B}_\text{F32}$ is split into two low-precision elements in $\mathbf{A}_\text{low}, \mathbf{B}_\text{low}$ and $\Delta\mathbf{A}_\text{low}, \Delta\mathbf{B}_\text{low}$, respectively.
Then the single-precision matrix multiplication is computed approximately in Eq. (\ref{eq:tcec-last}) using Tensor Cores, where the multiplication and addition are performed on specialized arithmetic units in a precision that is equivalent to FP32.
Furthermore, this method uses FP32 SIMT Cores for addition with RN (Round to Nearest, ties to even) mode for $\mathbf{A}_\text{low} \cdot \mathbf{B}_\text{low}$ to avoid the RZ (Round toward Zero) rounding inside Tensor Cores.
We show the matrix-matrix multiplication implementations and their supporting input accuracy in Table \ref{tab:gemm-implementation-table}.
When we use TF32 for the input type of Tensor Cores, we can emulate the single-precision matrix multiplication in both the mantissa and exponent (TF32TCEC).
On the other hand, when we use FP16 for the low-precision, the supported exponent range of the input matrices is limited (FP16TCEC).
However, the theoretical peak performance of FP16TCEC is higher than TF32TCEC, as shown in the table.
Therefore, there is a trade-off between the supported exponent range and the throughput.
We can achieve a higher throughput using FP16TCEC without loss of accuracy if the elements of the input matrices are in the supported representation range.
Furthermore, we can use these SGEMM emulation methods for a single-precision tensor contraction.
In the TTGT algorithm, the input tensors are reshaped to matrices, and the contraction is computed as matrix multiplication.
Thus, improving the throughput of GEMM leads to improving the throughput of tensor contraction.

\begin{table}[t]
\centering
\begin{tabular}{cc|cc}
                      &    & \multicolumn{2}{c}{Input type of Tensor Core}                                                                                                                                                              \\
                      &    & \multicolumn{1}{c|}{TF32}                                                                            & FP16                                                                                  \\
                      \hline
\multirow{4}{*}{\begin{tabular}[c]{@{}c@{}}Error \\ correction\end{tabular}} & Yes & \multicolumn{1}{c|}{\begin{tabular}[c]{@{}c@{}}{\bf TF32TCEC}\\ Exponent:FP32, Mantissa:FP32\\ {[}52 TFlop/s{]}\end{tabular}}  & \begin{tabular}[c]{@{}c@{}}{\bf FP16TCEC} \\Exponent:FP16, Mantissa:FP32\\ {[}104 TFlop/s{]}\end{tabular} \\ \cline{2-4} 
                      & No & \multicolumn{1}{c|}{\begin{tabular}[c]{@{}c@{}}{\bf TF32TC} \\Exponent:FP32, Mantissa:FP16\\ {[}156 TFlop/s{]}\end{tabular}} & \begin{tabular}[c]{@{}c@{}}{\bf FP16TC} \\Exponent:FP16, Mantissa:FP16\\ {[}312 TFlop/s{]}\end{tabular}      
\end{tabular}
\caption{The comparison of GEMM implementations using Tensor Cores on NVIDIA A100 GPU. Each throughput represents theoretical peak performance and is calculated by an assumption that the Tensor Core instruction is issued every clock.}

\label{tab:gemm-implementation-table}
\end{table}

\subsection{Quantum circuit simulation and tensor network contraction}
A quantum circuit consists of quantum gates and wires, as shown in Fig. \ref{fig:qc-intro} (a), similar to a classical logic circuit.
A quantum state of $n$ qubits is represented as a normalized complex vector of length $2^n$.
For instance, a typical $1$-qubit state called \textit{computational basis states} is represented as $\ket{0}:=(1, 0)^\mathsf{T}$ and $\ket{1}:=(0, 1)^\mathsf{T}$.
The quantum gate for $k$ qubits is represented as a $2^k \times 2^k$ unitary matrix.
For instance, the Hadamard gate, which is one of the single-qubit gates, is represented as $\mathbf{H}=\begin{bmatrix}1 & 1 \\ 1 & -1\end{bmatrix}/\sqrt{2}$, and T gate is $\mathbf{T}=\begin{bmatrix} 1 & 0 \\ 0 & \text{exp}(i\pi/4)\end{bmatrix}$, where $i$ is the imaginary unit.
The change of the quantum states by applying a quantum gate is represented as a multiplication of the quantum gate matrix and the state vector of the quantum states.
For instance, applying the Hadamard gate to the quantum state $\ket{0}$, we obtain the new state as follows:
$$
    \ket\psi = \mathbf{H}\ket{0}=\frac{1}{\sqrt{2}}
\left(
\begin{array}{c}
1 \\
1 \\
\end{array}
\right).
$$
The quantum state is a superposition state, from which we can not extract information directly.
Instead, we conduct \textit{measurements} to get the information of the quantum state indirectly.
In the case of an $n$-qubits system, a quantum state $\ket{\phi}$ is a superposition of $2^n$ states, $\ket{00\cdots 00}, \ket{00\cdots 01}, \cdots, \ket{11\cdots 11}$ in the computational basis, and represented as follows:
$$
\ket{\phi} = \sum_{j=00\cdots 00}^{11\cdots 11} \alpha_j \ket{j},
$$
where $\alpha_j$ is called the \textit{amplitude} of the state $\ket{j}$ and $\sum_{j=00\cdots 00}^{11\cdots 11} |\alpha_j|^2 = 1$.
When measuring the state $\ket{\phi}$, we obtain an $n$-length bitstring $j$ with a probability $|\alpha_j|^2$.
We conduct the measurements many times and obtain the distribution of the output bitstrings to extract the information of the quantum state.
In contrast, the quantum circuit simulation directly calculates the amplitude of given bitstrings.
However, calculating the amplitude of a large qubit system and circuit generally requires large computational costs and memory.


\begin{figure}
    \centering
    \includegraphics[width=0.7\linewidth]{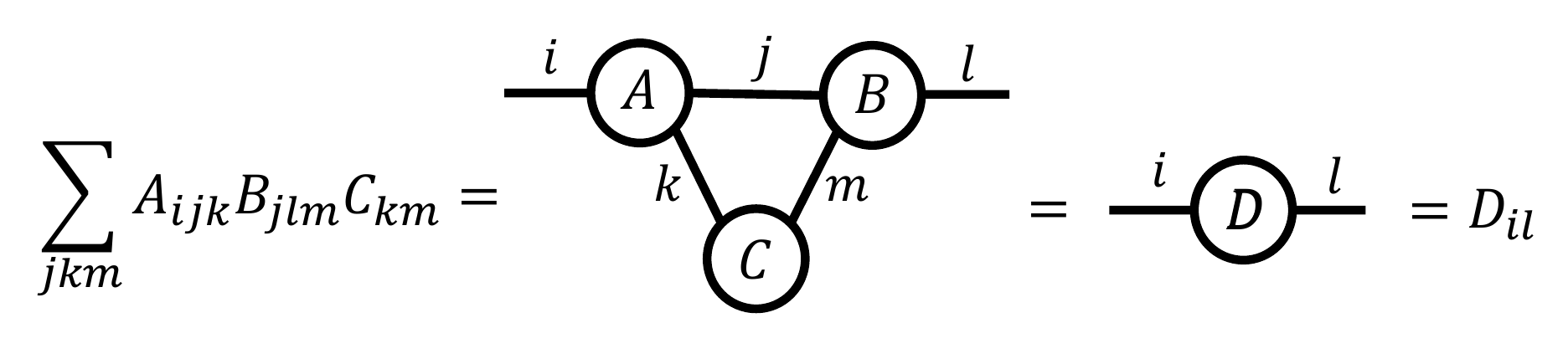}
    \caption{An example of the tensor network diagram.}
    \label{fig:TN}
\end{figure}

One of the methods to compute the amplitude is tensor network contraction.
A simple example of tensor network formalism is shown in Fig. \ref{fig:TN}. A rank-$r$ tensor, which is an element of $\mathbb{C}^{d_1\times\cdots\times d_r}$ with indices of dimension $d_i$, is represented as a node with $r$ edges. The connection of the edges in a network corresponds to Einstein's summation over the corresponding index.
A quantum circuit can be represented as a tensor network.
The initial state $\ket{00 \cdots 00}$ and the measuring state $\ket{x}$ are a set of rank-$1$ tensors, $n$-qubits gates are rank-$2n$ tensors, and the wires represent contraction.
The amplitude is calculated by the contraction of the whole tensor network.
Therefore, the primary workload of the quantum circuit simulation by the tensor network contraction is GEMM.

The time complexity of a tensor network contraction strongly depends on its contraction order or \textit{contraction path}.
In the example in Fig. \ref{fig:TN}, calculating in order $$
\sum_{jkm=1}^dA_{ijk}B_{jlm}C_{km}=\sum_{km}C_{km}\left(\sum_{j}A_{ijk}B_{jlm}\right)
$$
requires computational complexity $\mathcal{O}(d^5)$.
On the other hand,
$$
\sum_{jkm=1}^dA_{ijk}B_{jlm}C_{km}=\sum_{j}\left(\sum_{km}A_{ijk}B_{jlm}C_{km}\right)
$$
requires only $\mathcal{O}(d^4)$.
In general, finding the optimal contraction path is NP-complete\cite{chi-chung1997Optimizing} and even contracting the entire tensor network is \#P-complete\cite{schuch2007Computational}.
Therefore, there is no evidence that quantum circuit simulators by the tensor network contraction method perform well.
However, in practice, by using advanced techniques such as hyperparameter optimization and hypergraph partitioning to find a near-optimal contraction order and by massively distributing the computation, the tensor network methods outperform the state vector method, especially for systems with large qubits and shallow circuits\cite{huang2021Efficient, gray_hyper-optimized_2021, liu_closing_2021,pan2022Solving}.
For instance, \cite{liu_closing_2021} performed the Sycamore simulation in only 304 seconds by carefully selecting the contraction path and parallelizing on tens of millions of CPU cores.

\begin{figure}[t]
    \centering
    \includegraphics[width=0.9\linewidth]{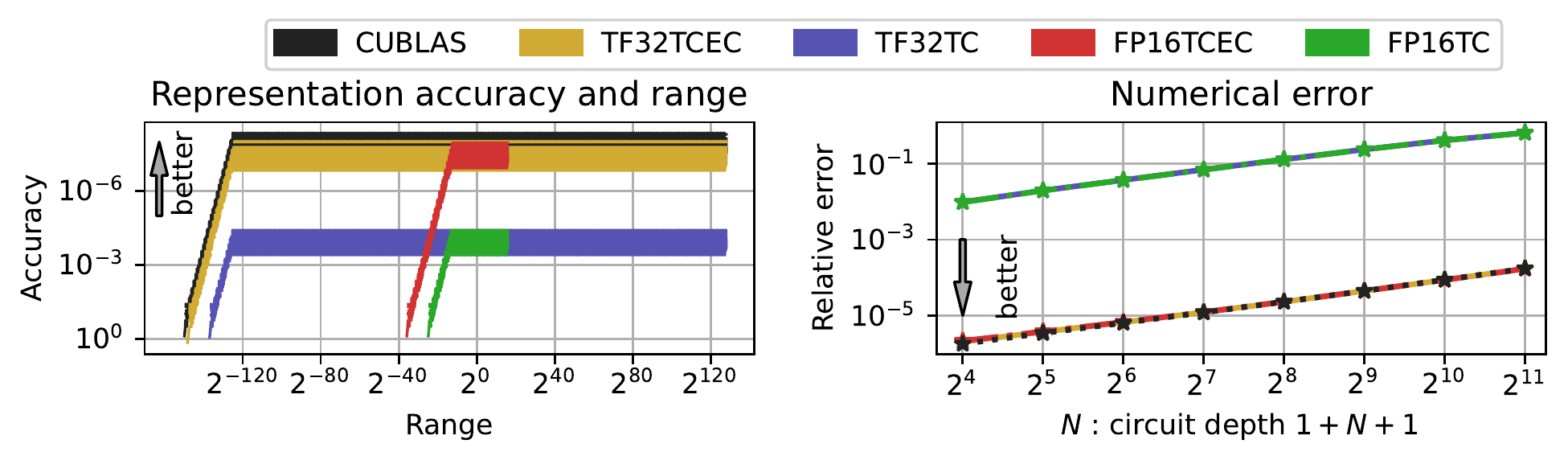}
    \caption{The evaluation of numerical error of $4\times 4$ Random Quantum Circuit simulation.}
    \label{fig:4x4-fidelity}
\end{figure}
The computation of tensor network contraction for quantum circuit simulation is performed in single precision.
For the computation of an $n$-qubit circuit simulation, it is required to be able to represent $2^{-(n-1)/2}$ by a floating point value at least to represent the resulting amplitude.
Therefore, FP16, which only has 5 bits of the exponent, is insufficient.
From the perspective of mantissa accuracy, we also need single precision.
We show the numerical accuracy of a quantum circuit simulation in Fig. \ref{fig:4x4-fidelity}, where the qubits are arranged in a $4 \times 4$ rectangular lattice, and the quantum circuit is RQC explained later in Sec. \ref{sec:rqc}.
Since the number of qubits $4 \times 4$ is small enough for using FP16 in terms of the exponent range, we can ignore the underflow error and evaluate only the effect of mantissa accuracy.
As we can see in the graph, the numerical errors of TF32TC and FP16TC are larger than the others.
This results in worse simulation accuracy (fidelity) in Fig. \ref{fig:intro-top}, even for relatively shallow circuits.
Therefore, using low-mantissa-length arithmetics without error correction is unsuitable for quantum circuit simulation.

\section{SGEMM emulation library on Tensor Cores}
\label{sec:sgemm-library}
We have implemented an SGEMM emulation library, cuMpSGEMM, that can be used in existing applications that call the NVIDIA cuBLAS SGEMM without modifying the source code of the applications.
The implementation is made from scratch using NVIDIA WMMA API (Tensor Core device API), and WMMA API extension library \cite{ootomo_reducing_2023}.
This library supports single-precision real and complex GEMM (SGEMM/CGEMM) and their batched variants.
We extend the SGEMM emulation to CGEMM by decomposing the real part and imaginary part of the input matrices and computing it as 4 SGEMMs as follows:
\begin{align*}
    \mathbf{C}_\text{F32}^\text{complex} \leftarrow& \mathbf{A}_\text{F32}^\text{complex} \cdot \mathbf{B}_\text{F32}^\text{complex} \\
    =& \left(\mathbf{A}_\text{F32}^\text{real} \cdot \mathbf{B}_\text{F32}^\text{real} - \mathbf{A}_\text{F32}^\text{imag} \cdot \mathbf{B}_\text{F32}^\text{imag}\right)
    +\left(\mathbf{A}_\text{F32}^\text{real} \cdot \mathbf{B}_\text{F32}^\text{imag} + \mathbf{A}_\text{F32}^\text{real} \cdot \mathbf{B}_\text{F32}^\text{imag}\right)i.
\end{align*}
This library intercepts the function calls to cuBLAS SGEMM functions and executes the SGEMM emulation functions instead of the original functions.
Therefore, we can use the SGEMM emulation without changing the source code of the target application if the application uses the cuBLAS dynamic library.
All we need to do is to build the library, set an environmental variable {\tt LD\_PRELOAD}, and execute the target application as usual.
We have confirmed that this library can intercept cuBLAS calls in PyTorch \cite{paszke_pytorch_2019}, CuPy \cite{okuta_cupy_2017}, and our custom applications just by following these steps.
The different implementations shown in Table \ref{tab:gemm-implementation-table} can be selected by defining an environment variable.

\begin{figure}[t]
    \centering
    \includegraphics[width=0.85\linewidth]{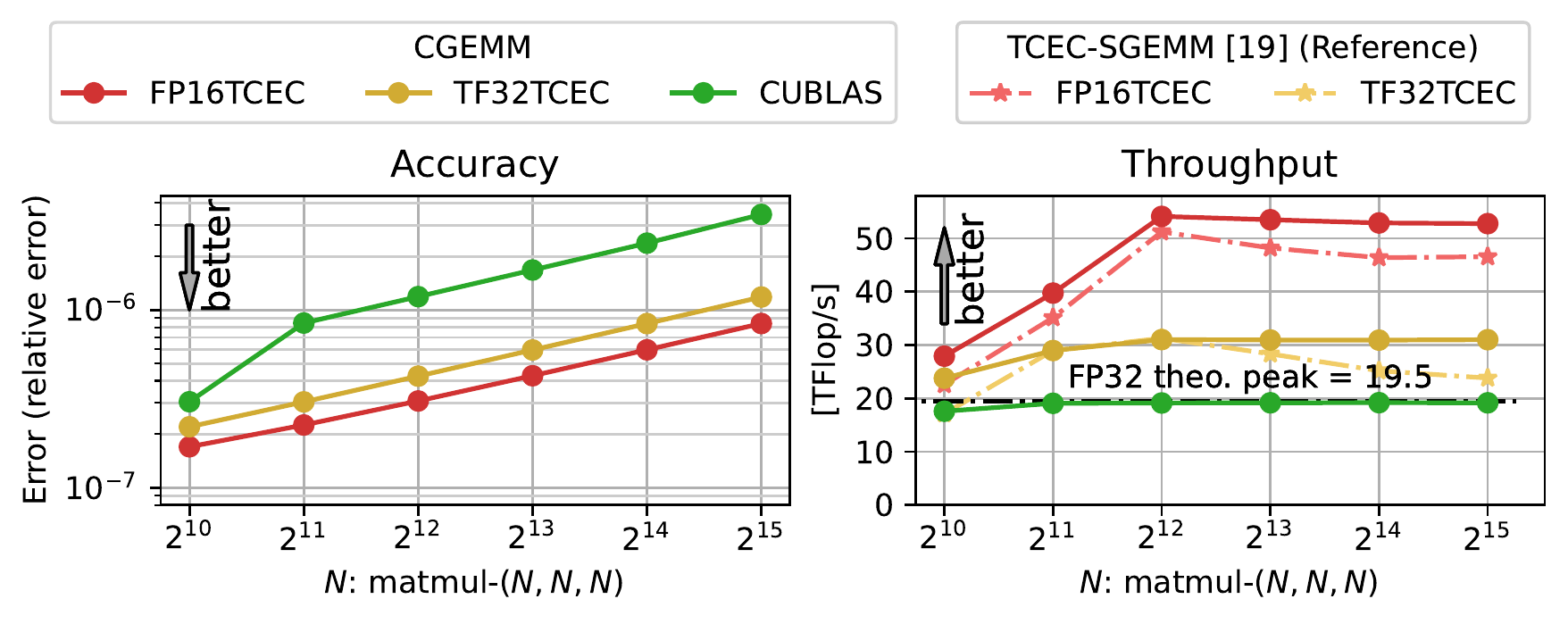}
    \caption{The accuracy and throughput of the CGEMM implementation on Tensor Cores using the error correction method.}
    \label{fig:gemm-eval}
\end{figure}
We perform a unit test for our CGEMM implementation as shown in Fig. \ref{fig:gemm-eval}.
We denote the shape of GEMM as $(m, n, k)$, which is the multiplication of $m \times k$ and $k \times n$ matrices.
To measure the accuracy, we calculate a relative error of a single-precision complex matrix-matrix multiplication $\mathbf{C}_\text{F32} \leftarrow \mathbf{A}_\text{F32} \cdot \mathbf{B}_\text{F32}$ as follows:
\begin{equation}
    \text{Relative Error} = ||\mathbf{C}_\text{F32} - \mathbf{C}_\text{F64}||_F/||\mathbf{C}_\text{F64}||_F,
\end{equation}
where $\mathbf{C}_\text{F64}$ is the result in FP64, and each element of $\mathbf{A}_\text{F32}$ and $\mathbf{B}_\text{F32}$ is chosen from a uniform distribution $(-1, 1)$.
Since the values are in the order of $10^{-7} \sim 10^{-6}$ for each matrix size, FP32 is sufficient, but FP16 is not.
To evaluate the throughput, we measure the computing time and calculate the throughput.
The maximum throughput of FP16TCEC and TF32TCEC are 54.2 [TFlop/s] and 31.0 [TFlop/s], respectively. This performance is almost identical to the TCEC-SGEMM implementation using NVIDIA CUTLASS \cite{ootomo_recovering_2022}.

\section{Automatic precision selection}
\begin{figure}[t]
    \centering
    \includegraphics[width=\linewidth]{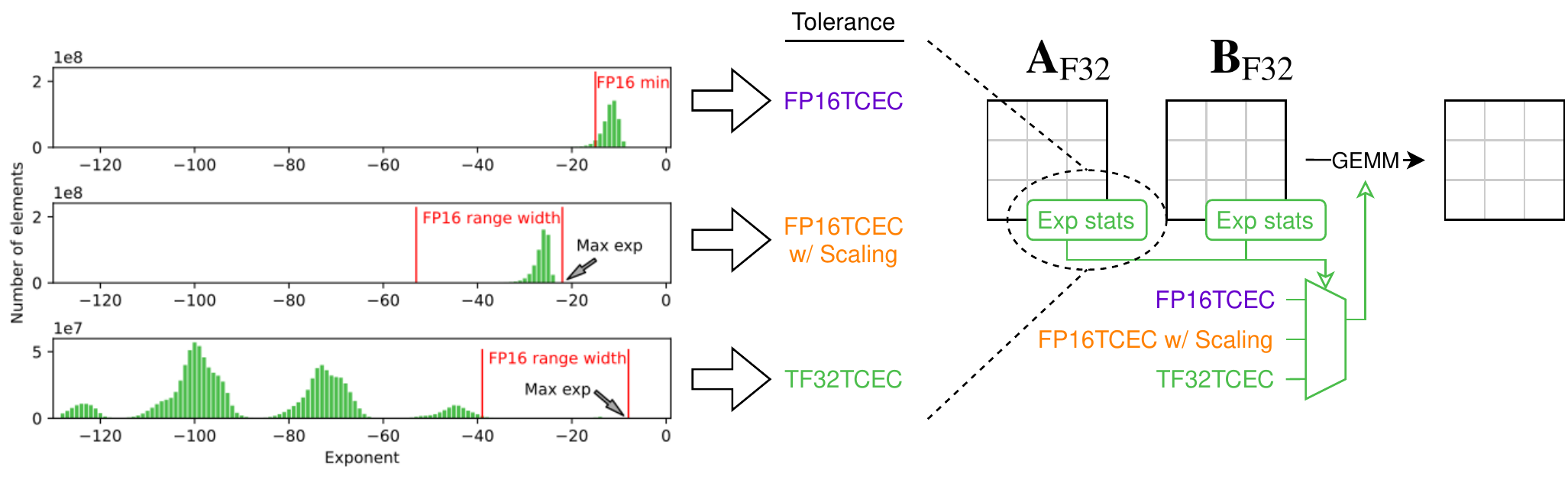}
    \caption{The overview of the automatic kernel selection by the exponent statistics of the input matrices.}
    \label{fig:exp-stats-example}
\end{figure}

Although FP16TCEC supports a limited exponent range compared to TF32TCEC, it is faster than TF32TCEC.
Therefore, we can improve SGEMM throughput by selecting FP16TCEC when its accuracy loss is permissible.
However, it is generally impossible to predetermine whether FP16TCEC is tolerable before the actual computation.
To check the tolerance, we take the statistics of the exponent distribution of the input matrices before the GEMM operation, as shown in Fig. \ref{fig:exp-stats-example}, and select the GEMM computing mode dynamically.
In addition to FP16TCEC and TF32TCEC, we can select {\bf FP16TCEC w/ scaling} mode.
This mode can be used when the required exponent range is sufficient for FP16, even when the values themselves are small enough to underflow in FP16.
This requires an additional overhead for scaling all elements in the input and output matrices.

The challenge in implementing the automatic precision selection method is to get the statistics correct while keeping its computational overhead negligibly small relative to the GEMM computing time.
To achieve this, we adopt two strategies as follows:
\begin{itemize}
    \item We check all elements in the matrix to take the statistics.
    Although we can reduce the overhead by sampling only part of the elements, this can lead to overflow in some cases.
    We check all elements to prevent overflow and compute the tensor contraction correctly.
    \item We do not transfer the statistics data from GPU to CPU to minimize the overhead.
    Although we need the exponent statistics to control which GEMM kernel we use, it results in an additional overhead for synchronizing CUDA kernels and sending the statistics data on the device memory to host memory.
    To reduce the overhead, we preemptively launch all CUDA kernels and kill some if they are not necessary instead of controlling the kernels to be launched.
\end{itemize}
We explain the detail of the two components above in the following sections.

\subsection{Exponent statistics and computing mode selection rule}
\label{sec:mode-selection-rule}
To determine that a given tensor can tolerate FP16TCEC or FP16TCEC w/ scaling, we take the exponent statistics of each element in two stages.
\begin{enumerate}[label=Stage \arabic*:, leftmargin=*]
    \item Obtain the number of elements that are larger than the minimum value of FP16 ($N_1$) and the max value of the exponent ($e_\text{max}$).
    When $N_1$ is larger than the underflow admissibility threshold, we mark the matrix that can tolerate FP16TCEC and skip the next stage.
    \item Obtain the number of elements within the shifted FP16 exponent range ($N_2$), where the range is shifted so that the maximum exponent becomes $14$, which is the maximum exponent value of FP16.
    When $N_2$ is larger than the threshold, we mark the matrix that can tolerate FP16TCEC w/ scaling; otherwise, we use TF32TCEC.
\end{enumerate}
The threshold is shared in the two stages above and can control the accuracy of the tensor contraction.
We check the tolerance for both input matrices.
When either input matrix can not tolerate FP16TCEC and FP16TCEC w/ scaling, we use TF32TCEC for the GEMM operation.
When either of the input matrices can not tolerate FP16TCEC, we use FP16TCEC w/ scaling.
Otherwise, we use FP16TCEC.
Note that we only check the underflow since the absolute values of all real and imaginary values of tensor elements in the quantum circuit simulation are smaller or equal to $1$. 

\subsection{Dynamic kernel selection}
\begin{figure}[t]
    \centering
    \includegraphics[width=\linewidth]{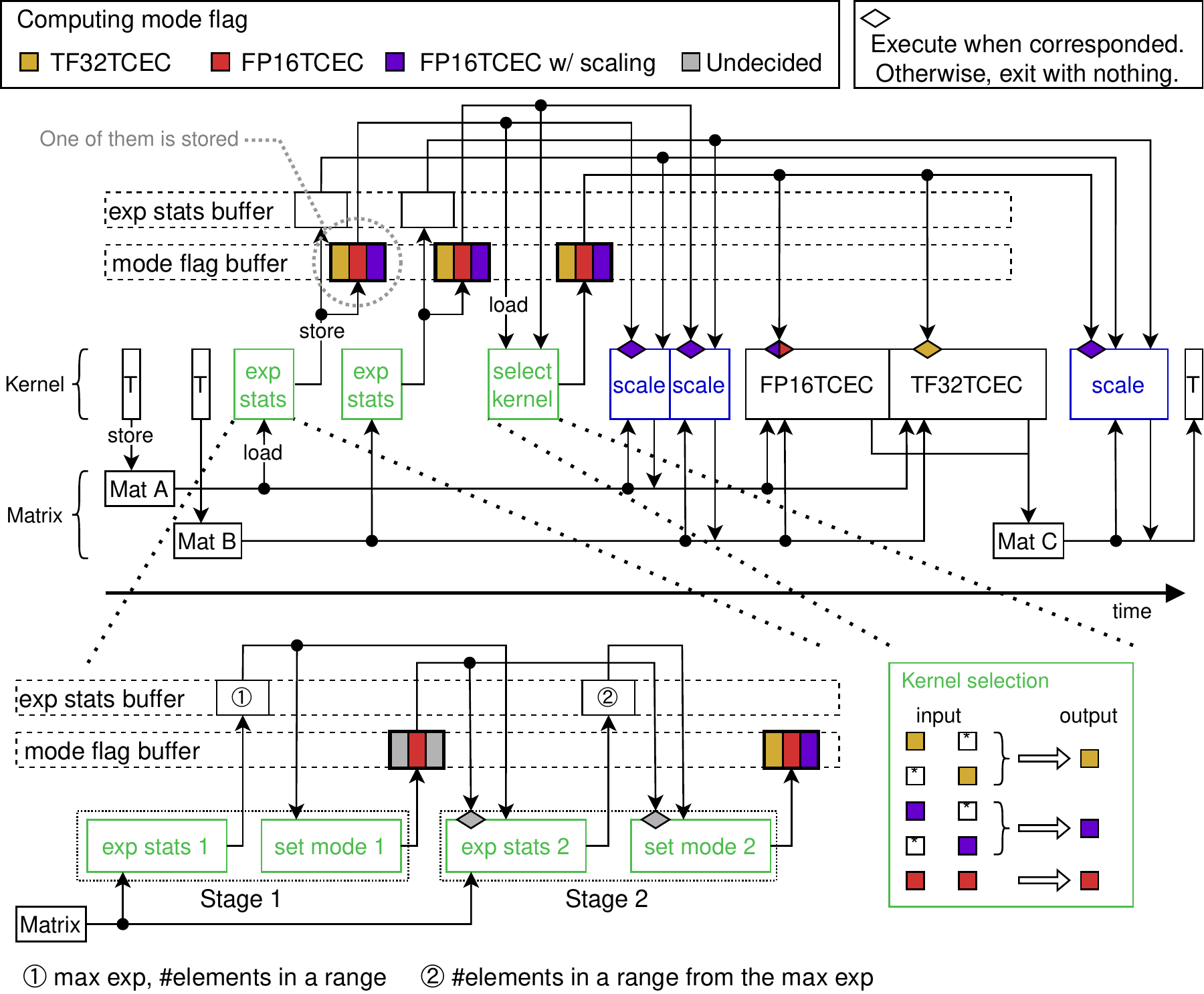}
    \caption{
    The dynamic kernel selection mechanism to compute a tensor contraction (TTGT) using the {\tt mode flag buffer} on the device memory.
    }
    \label{fig:ttgt-w-exp-stats}
\end{figure}
We select the GEMM kernel function depending on the result of the exponent statistics of two input matrices.
The simplest way to realize this is to offload the statistics to the host and launch the selected GEMM kernel function.
However, this requires GPU-to-host data transfer and may result in throughput degradation.
Although NVIDIA provides the Dynamic Parallelism API for such situations that can launch a kernel function from another, we can not use this API since there is a limitation related to the dynamic shared memory size configuration we use.
Instead, we use another method: we launch all CUDA kernel functions possibly used and kill some if they are not.
For instance, we launch both TF32TCEC and FP16TCEC kernel functions, but one of them exits at the beginning of the execution by checking the computing mode flag on the device memory.
The entire flow of execution is shown in Fig. \ref{fig:ttgt-w-exp-stats}.
We use two buffers on the device memory to control the kernel execution, {\tt exp stats buffer} and {\tt mode flag buffer}.
The kernel function for taking the exponent statistics stores $N_1, N_2$ and $e_\text{max}$ on {\tt exp stats buffer} and the tolerance on {\tt mode flag buffer}.
Based on the {\tt mode flag}s of input matrices, the GEMM mode selection kernel ({\tt select kernel}) selects the computing mode by the selection rule in \ref{sec:mode-selection-rule}.
When the mode is FP16TCEC w/ scaling, the scaling kernel ({\tt scale}) scales the matrix elements so that the maximum exponent value becomes $14$, and after computing the GEMM operation in FP16TCEC, we scale the resulting matrix to balance out.
Note that we do not restore the scaled input matrices since they are not reused in the quantum circuit simulation\footnote{The library itself has an optional functionality to restore the scaled input matrices for general purpose.}.

\subsection{The overhead of the exponent statistics}
\begin{figure}[t]
    \centering
    \includegraphics[width=\linewidth]{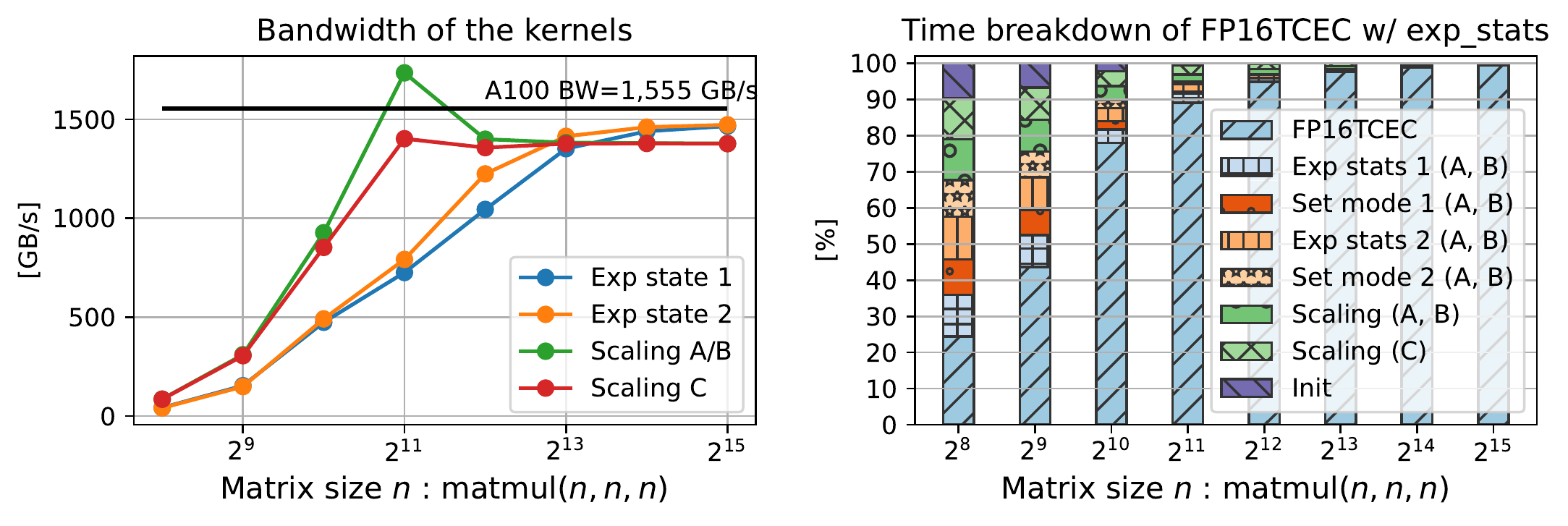}
    \caption{
    {\bf Left: } The bandwidth efficiency of the exponent statistics and scaling kernel function.
    {\bf Right: } The breakdown of the automatic precision selection in the case where FP16TCEC is selected.}
    \label{fig:exp-stats-overhead}
\end{figure}
The operations for taking the exponent statistics and scaling elements are memory bandwidth intensive.
We have measured their throughput efficiency shown on the left side of Fig. \ref{fig:exp-stats-overhead}.
When the matrix size is large, they achieve more than 90 \% of the theoretical peak throughput.
In the case of $n=2^{11}$, the throughput of the {\tt Scaling A/B} kernel function outperforms the theoretical bandwidth since the data is on the L2 cache, since it is loaded in the {\tt Exp stats 2} kernel function executed just prior to it.
We have also investigated the time breakdown of the automatic precision selection shown on the right of Fig. \ref{fig:ttgt-w-exp-stats}.
When the matrix size is small, the overheads of the automatic precision selection and the scaling operation are not negligible.
Therefore, we use the following rule for automatic precision selection when the matrix size is large:
\begin{itemize}
    \item When $m, n, k \geq 2048$: Use the automatic precision selection.
    \item Otherwise and when $m, n, k \geq 512$: Use TF32TCEC.
    \item Otherwise: Use cuBLAS.
\end{itemize}

\section{Experiment}
\subsection{Preparation}
\subsubsection{Quantum circuit simulation on GPU}
We use the Python library TensorNetwork \cite{roberts2019TensorNetwork}, and quimb \cite{gray_quimb_2018} for the quantum circuit simulation using the tensor network contraction, cotengra \cite{gray_hyper-optimized_2021}, kahypar \cite{schlag2022HighQuality} and  opt\_einsum \cite{g.a.smith2018Opt} for path optimization, and CuPy \cite{okuta_cupy_2017} for the computational backend on GPUs.
Although there are several quantum circuit simulators such as NVIDIA cuTensorNet\footnote{\url{https://docs.nvidia.com/cuda/cuquantum/cutensornet/index.html}}, it is not possible to perform a fair comparison against them since their method to avoid explicit transposing on device memory is not open-source.
While our method focuses on improving the CGEMM throughput after transposing input tensors, cuTensor, which is used in cuTensorNet for the tensor contraction, focuses on reducing data movement by avoiding explicit transposing on device memory, etc.
Therefore, technically, we can apply our method to cuTensorNet if it is open-source.
We have confirmed experimentally that cuTensorNet is faster than our methods depending on the problem sizes and contraction paths since the GEMM and transposition time ratio vary.

\subsubsection{Fast implementation for irregular shaped GEMM}
\label{sec:irregular-shaped-gemm}
In quantum circuit simulation, the irregular shape of GEMMs, for instance, $(2, 2^N, 2)$ and $(2^N, 2, 2)$ for $N \geq 10$, are computed many times for the contraction of a single-qubit gate tensor.
However, the cuBLAS CGEMM function for these shapes is not optimized.
The memory bandwidth bounds the throughput of the GEMMs for these shapes.
We have implemented specialized CGEMM kernel functions for these shapes and achieved up to $90$\% of the theoretical device memory bandwidth on A100 GPU, while that of cuBLAS is $10$\%.
We enable this kernel function in all experiments in all computing modes, including the baseline CUBLAS mode.

\subsection{Exploratory experiment}
\begin{figure}[t]
    \centering
    \includegraphics[width=\textwidth]{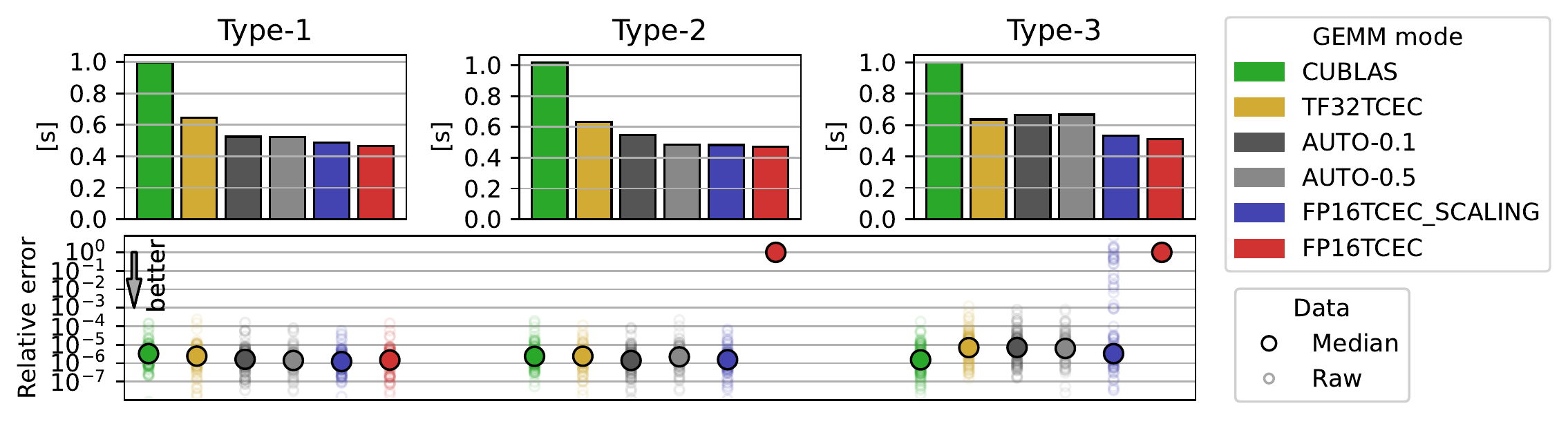}
    \caption{The computing time (top) and accuracy (bottom) of a random tensor network contraction for different types of element initializations.}
    \label{fig:randtn-result}
\end{figure}
We check the behavior of the automatic precision selection on a tensor network contraction which is randomly constructed.
The tensor network consists of 10 nodes with $2 \sim 4$ degrees.
Each dimension of the tensors is 128, and all elements are initialized as follows.
\begin{enumerate}[label=Type-\arabic*:, leftmargin=*]
    \item With standard distribution $\mathcal{N}(0, 10^{-4})$.
    All computing modes can compute with high accuracy.
    \item After Type-1 initialization, all elements are scaled $10^{-6}$.
    FP16TCEC can not compute with high accuracy without scaling.
    \item After Type-2 initialization, set $10\sim 20$ elements $1$.
    FP16TCEC can not compute with high accuracy, even with scaling.
\end{enumerate}
The accuracy is the error relative to FP64 computation.
We denote our method as AUTO-$t$, where $t$ is the underflow tolerance threshold.
For instance, when $t=0$, the mode that can avoid all underflows is selected.
We show the computing time and accuracy in Fig. \ref{fig:randtn-result}.
Throughout all types, the AUTO modes select the proper computing mode to avoid underflow automatically, and their computing accuracy is close to the same level as the baseline (CUBLAS).
In Type-2, the computing time of AUTO-$0.5$ is shorter than AUTO-$0.1$ since it selects different computing modes while maintaining accuracy.
That implies that the underflow tolerance is being used in some cases.
However, it is not feasible to find an appropriate tolerance value $t$ since it is also related to the positional distribution of the input matrix elements, such as sparsity.
In Type-2 and 3, the accuracy of FP16TCEC underflows to zero.
Although the median error of TF32TCEC, AUTO, and FP16TCEC w/ scaling is larger than the baseline by about 2 bits of mantissa, we consider that the cause is computation order, and the accuracy is considered sufficiently single-precision.
However, each result of FP16TCEC w/ scaling sometimes underflows to zero or has low accuracy.
In summary, we have confirmed that the AUTO mode automatically selects high throughput computing mode while achieving the same level of accuracy as the baseline.

\subsection{Random Quantum Circuit simulation}
\label{sec:rqc}

We evaluate the automatic precision selection on the Random Circuit Sampling (RCS) problem\cite{boixo_characterizing_2018}, which is the task to demonstrate the so-called \textit{quantum advantage}\cite{preskill2012Quantum}.
In the RCS problem, we sample the output bitstrings of a random quantum circuit (RQC) $U$.
The bitstring $x$ follows the distribution:
\begin{equation}
    \label{eq:prob_dist}
    p_U(x):=|\braket{x|U|0}|^2,
\end{equation}
where $p_U(x)$ is the appearance probability of $x$.
On a quantum device, the sampling can be accomplished by preparing an initial zero state, applying a unitary circuit $U$, and measuring each qubit on a computational basis.
On the other hand, on a classical computer, we first decide a (random) bitstring and compute its appearance probability through Eq. \ref{eq:prob_dist}.
From the computed probability, we decide whether we accept the bitstring as the output of the quantum circuit or not.
Since the computation of one amplitude is independent of other amplitude computations, the throughput scales linearly with computing resources.
Generally, the sampling cost on a classical computer can be much higher than on a quantum device since we reject much more bitstrings than we accept.
However, in the case of RCS, the \textit{frugal reject sampling}\cite{markov_quantum_2018} technique can be used to sample them on a classical computer at about the same computational cost as on a quantum device by reducing the number of rejected bitstrings.
Therefore, many studies focus on improving the throughput of computing one amplitude.

We evaluate our method on the rectangular lattice RQC defined in \cite{boixo_characterizing_2018} and Sycamore circuit \cite{arute_quantum_2019-2}.
As mentioned above, all quantum circuits and measurements are represented as tensor networks, and one amplitude is obtained by contracting them.
Since the computational cost of the Sycamore circuit is high, we typically divide it into slices and merge the result of their contraction.
We measure the computing time for one amplitude for the rectangular lattice RQC and one slice for the Sycamore circuit.
The computation is conducted $10$ times for different output bitstrings, and we show their median.

\subsubsection{Rectangular lattice RQC}
\begin{figure}[t]
    \centering
    \includegraphics[width=\textwidth]{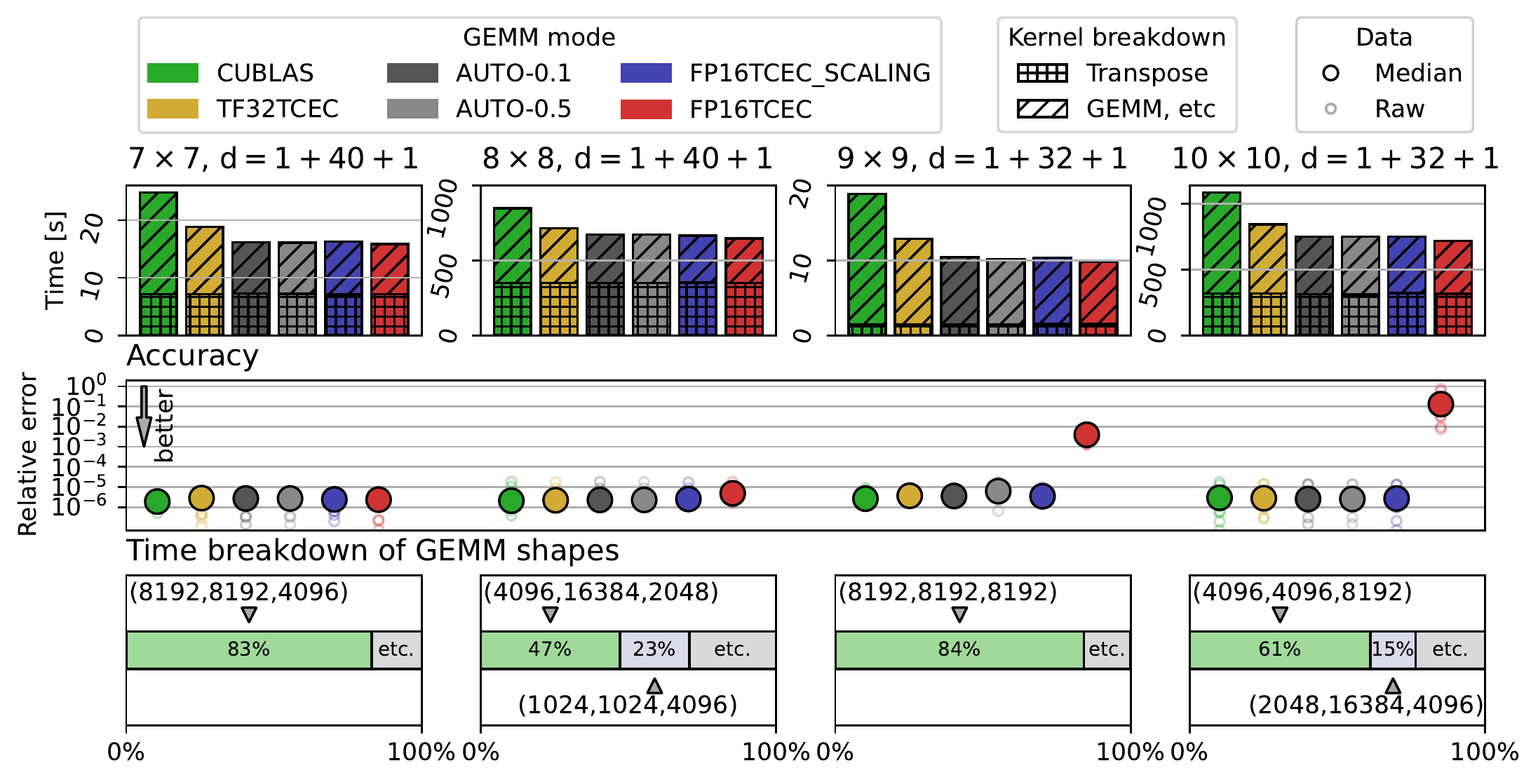}
    \caption{
    The computing time (top) and accuracy (center) of obtaining one amplitude for each RQC and the GEMM shape time breakdown in the CUBLAS mode (bottom) of the rectangular lattice RQCs.
    }
    \label{fig:rqc-result}
\end{figure}
The rectangular lattice RQC has $m \times n$ qubits arranged in a lattice, and we apply a specific set of quantum gates.
Furthermore, the quantum gates are also arranged in a rectangular lattice, and the number of the rectangular lattice layers is called ``depth".
We denote the depth $d=1 + X + 1$ since we apply the Hadamard gate for all qubits in the first ($1+$) and last (+1) stages of the circuit.
We have evaluated the automatic precision selection on four kinds of rectangular lattice RQCs, as shown in Fig. \ref{fig:rqc-result}.
Throughout all circuits, the AUTO modes avoid the underflow automatically and achieve the same level of accuracy as the baseline (CUBLAS) while achieving higher throughput than TF32TCEC.
The AUTO-$0.5$ mode has achieved up to $1.86$ times higher throughput than the baseline in the $9 \times 9$ quantum circuit, which has a small computing time ratio for the tensor transposition.
The TF32TCEC mode has not been selected in the AUTO modes since it has sufficed to use FP16TCEC w/ scaling in all cases.
Therefore, the throughput of the AUTO modes is almost the same as FP16TCEC w/ scaling mode.
Although the throughput of FP16TCEC is also higher than TF32TCEC, accuracy loss occurs due to the underflow when the qubit size is large.
We have also investigated the computing time breakdown for GEMM shapes in each quantum circuit.
In all circuits, a large shape of GEMM dominates the whole GEMM computing time.
In this case, we can improve the efficiency of our method since the SGEMM emulation methods achieve higher throughput, and the exponent statistics have a relatively lower overhead.

\subsubsection{Sycamore}
\begin{figure}[t]
    \centering
    \includegraphics[width=\linewidth]{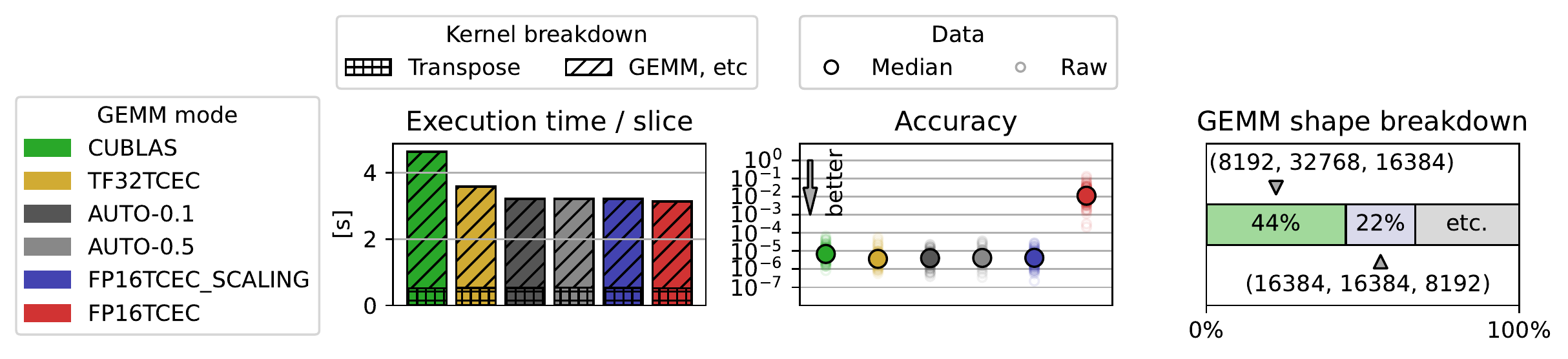}
    \caption{The computing time (left), accuracy (center), and GEMM shape time breakdown in the CUBLAS mode (right) of Google Sycamore sampling simulation.}
    \label{fig:sycamore-eval}
\end{figure}
In the Sycamore circuits, 53 qubits are arranged in 2-D, and we apply $(1+20+1)$-depth quantum gates layers.
The evaluation result of the Sycamore circuit is shown in Fig. \ref{fig:sycamore-eval}.
The AUTO modes have achieved $1.45$ times higher throughput than the baseline (CUBLAS) while achieving the same level of accuracy.
In this quantum circuit, the large shape GEMMs dominate the whole GEMM computing time, and we can efficiently improve the throughput of the simulation.
Although the underflow ratio of the elements in the large GEMMs has differed from the output strings, FP16TCEC w/ scaling is selected in all computations.
As we mentioned above, since the throughput of the RCS is proportional to the throughput of one amplitude computation, we believe our method improves the whole Sycamore circuit simulation to the same extent.

\section{Conclusion}
We improve the throughput of the quantum circuit simulation using the SGEMM emulation method on Tensor Cores and automatic precision selection.
Our method automatically selects the GEMM computing modes in the tensor contractions to improve the throughput while avoiding accuracy loss due to underflow.
We have achieved up to $1.86$ times throughput in $9 \times 9$ RQC and $1.45$ times in Google Sycamore circuits compared to the baseline simulation using cuBLAS CGEMM while keeping the accuracy. Furthermore, we have also achieved up to $1.27$ times in $9 \times 9$ RQC, $1.11$ times faster throughput in the Sycamore circuit than SGEMM emulation using TF32 Tensor Core, which clearly can improve the throughput.
Through this study, we show an example that Tensor Core, which is developed for machine learning, can be used for another HPC field of research.

\subsubsection*{Acknowledgements}
This work was partially supported by JSPS KAKENHI JP22H03598, JP21J14694, and JP20K03766.
This work was partially supported by "Joint Usage/Research Center for Interdisciplinary Large-scale Information Infrastructures" in Japan (Project ID: jh220022-NAHI).

\bibliography{refs, bibs_manabe}

\begin{thebibliography}{30}
\providecommand{\natexlab}[1]{#1}
\providecommand{\url}[1]{\texttt{#1}}
\expandafter\ifx\csname urlstyle\endcsname\relax
  \providecommand{\doi}[1]{doi: #1}\else
  \providecommand{\doi}{doi: \begingroup \urlstyle{rm}\Url}\fi

\bibitem[Arute et~al.(2019)Arute, Arya, et~al.]{arute_quantum_2019-2}
Frank Arute, Kunal Arya, et~al.
\newblock Quantum supremacy using a programmable superconducting processor.
\newblock \emph{Nature}, 574\penalty0 (7779):\penalty0 505--510, October 2019.
\newblock ISSN 1476-4687.
\newblock \href{https://doi.org/10.1038/s41586-019-1666-5}{\ttfamily\path{
  doi:10.1038/s41586-019-1666-5}}.
\newblock URL \url{https://www.nature.com/articles/s41586-019-1666-5}.
\newblock Number: 7779 Publisher: Nature Publishing Group.

\bibitem[Boixo et~al.(2018)Boixo, Isakov, Smelyanskiy, Babbush, Ding, Jiang,
  Bremner, Martinis, and Neven]{boixo_characterizing_2018}
Sergio Boixo, Sergei~V. Isakov, Vadim~N. Smelyanskiy, Ryan Babbush, Nan Ding,
  Zhang Jiang, Michael~J. Bremner, John~M. Martinis, and Hartmut Neven.
\newblock Characterizing quantum supremacy in near-term devices.
\newblock \emph{Nature Physics}, 14\penalty0 (6):\penalty0 595--600, June 2018.
\newblock ISSN 1745-2481.
\newblock \href{https://doi.org/10.1038/s41567-018-0124-x}{\ttfamily\path{
  doi:10.1038/s41567-018-0124-x}}.
\newblock URL \url{https://www.nature.com/articles/s41567-018-0124-x}.
\newblock Bandiera\_abtest: a Cg\_type: Nature Research Journals Number: 6
  Primary\_atype: Research Publisher: Nature Publishing Group Subject\_term:
  Quantum information;Quantum simulation Subject\_term\_id:
  quantum-information;quantum-simulation.

\bibitem[Chen et~al.(2018)Chen, Zhou, Xue, Yang, Guo, and
  Guo]{chen_64-qubit_2018}
Zhao-Yun Chen, Qi~Zhou, Cheng Xue, Xia Yang, Guang-Can Guo, and Guo-Ping Guo.
\newblock 64-qubit quantum circuit simulation.
\newblock \emph{Science Bulletin}, 63\penalty0 (15):\penalty0 964--971, August
  2018.
\newblock ISSN 2095-9273.
\newblock \href{https://doi.org/10.1016/j.scib.2018.06.007}{\ttfamily\path{
  doi:10.1016/j.scib.2018.06.007}}.
\newblock URL
  \url{https://www.sciencedirect.com/science/article/pii/S2095927318302809}.

\bibitem[{Chi-Chung} et~al.(1997){Chi-Chung}, Sadayappan, and
  Wenger]{chi-chung1997Optimizing}
Lam {Chi-Chung}, P.~Sadayappan, and Rephael Wenger.
\newblock On {{Optimizing}} a {{Class}} of {{Multi-Dimensional Loops}} with
  {{Reduction}} for {{Parallel Execution}}.
\newblock \emph{Parallel Processing Letters}, 07\penalty0 (02):\penalty0
  157--168, June 1997.
\newblock ISSN 0129-6264.
\newblock \href{https://doi.org/10.1142/S0129626497000176}{\ttfamily\path{
  doi:10.1142/S0129626497000176}}.

\bibitem[G.~A.~Smith and Gray(2018)]{g.a.smith2018Opt}
Daniel G.~A.~Smith and Johnnie Gray.
\newblock Opt\textbackslash\_einsum - {{A Python}} package for optimizing
  contraction order for einsum-like expressions.
\newblock \emph{Journal of Open Source Software}, 3\penalty0 (26):\penalty0
  753, June 2018.
\newblock ISSN 2475-9066.
\newblock \href{https://doi.org/10.21105/joss.00753}{\ttfamily\path{
  doi:10.21105/joss.00753}}.

\bibitem[Gray(2018)]{gray_quimb_2018}
Johnnie Gray.
\newblock quimb: {A} python package for quantum information and many-body
  calculations.
\newblock \emph{Journal of Open Source Software}, 3\penalty0 (29):\penalty0
  819, September 2018.
\newblock ISSN 2475-9066.
\newblock \href{https://doi.org/10.21105/joss.00819}{\ttfamily\path{
  doi:10.21105/joss.00819}}.
\newblock URL \url{https://joss.theoj.org/papers/10.21105/joss.00819}.

\bibitem[Gray and Kourtis(2021)]{gray_hyper-optimized_2021}
Johnnie Gray and Stefanos Kourtis.
\newblock Hyper-optimized tensor network contraction.
\newblock \emph{Quantum}, 5:\penalty0 410, March 2021.
\newblock ISSN 2521-327X.
\newblock \href{https://doi.org/10.22331/q-2021-03-15-410}{\ttfamily\path{
  doi:10.22331/q-2021-03-15-410}}.
\newblock URL \url{http://arxiv.org/abs/2002.01935}.
\newblock arXiv: 2002.01935.

\bibitem[Guerreschi et~al.(2020)Guerreschi, Hogaboam, Baruffa, and
  Sawaya]{guerreschi_intel_2020}
Gian~Giacomo Guerreschi, Justin Hogaboam, Fabio Baruffa, and Nicolas P.~D.
  Sawaya.
\newblock Intel {Quantum} {Simulator}: a cloud-ready high-performance simulator
  of quantum circuits.
\newblock \emph{Quantum Science and Technology}, 5\penalty0 (3):\penalty0
  034007, May 2020.
\newblock ISSN 2058-9565.
\newblock \href{https://doi.org/10.1088/2058-9565/ab8505}{\ttfamily\path{
  doi:10.1088/2058-9565/ab8505}}.
\newblock URL \url{https://doi.org/10.1088/2058-9565/ab8505}.
\newblock Publisher: IOP Publishing.

\bibitem[Huang et~al.(2021)Huang, Zhang, Newman, et~al.]{huang2021Efficient}
Cupjin Huang, Fang Zhang, Michael Newman, et~al.
\newblock Efficient parallelization of tensor network contraction for
  simulating quantum computation.
\newblock \emph{Nature Computational Science}, 1\penalty0 (9):\penalty0
  578--587, September 2021.
\newblock ISSN 2662-8457.
\newblock \href{https://doi.org/10.1038/s43588-021-00119-7}{\ttfamily\path{
  doi:10.1038/s43588-021-00119-7}}.

\bibitem[Huang et~al.(2018)Huang, Yu, and van~de
  Geijn]{huang_implementing_2018}
Jianyu Huang, Chenhan~D. Yu, and Robert~A. van~de Geijn.
\newblock Implementing {Strassen}'s {Algorithm} with {CUTLASS} on {NVIDIA}
  {Volta} {GPUs}.
\newblock \emph{arXiv:1808.07984 [cs]}, August 2018.
\newblock URL \url{http://arxiv.org/abs/1808.07984}.
\newblock arXiv: 1808.07984.

\bibitem[Jones et~al.(2019)Jones, Brown, Bush, and Benjamin]{jones_quest_2019}
Tyson Jones, Anna Brown, Ian Bush, and Simon~C. Benjamin.
\newblock {QuEST} and {High} {Performance} {Simulation} of {Quantum}
  {Computers}.
\newblock \emph{Scientific Reports}, 9\penalty0 (1):\penalty0 10736, July 2019.
\newblock ISSN 2045-2322.
\newblock \href{https://doi.org/10.1038/s41598-019-47174-9}{\ttfamily\path{
  doi:10.1038/s41598-019-47174-9}}.
\newblock URL \url{https://www.nature.com/articles/s41598-019-47174-9}.
\newblock Number: 1 Publisher: Nature Publishing Group.

\bibitem[Liang et~al.(2021)Liang, Xu, Deng, Yan, Hu, Zhang, Li, and
  Xie]{liang_fast_2021}
Ling Liang, Jianyu Xu, Lei Deng, Mingyu Yan, Xing Hu, Zheng Zhang, Guoqi Li,
  and Yuan Xie.
\newblock Fast {Search} of the {Optimal} {Contraction} {Sequence} in {Tensor}
  {Networks}.
\newblock \emph{IEEE Journal of Selected Topics in Signal Processing},
  15\penalty0 (3):\penalty0 574--586, April 2021.
\newblock ISSN 1941-0484.
\newblock \href{https://doi.org/10.1109/JSTSP.2021.3051231}{\ttfamily\path{
  doi:10.1109/JSTSP.2021.3051231}}.
\newblock Conference Name: IEEE Journal of Selected Topics in Signal
  Processing.

\bibitem[Liu et~al.(2021)Liu, Liu, Li, Fu, Yang, Song, Zhao, Wang, Peng, Chen,
  Guo, Huang, Wu, and Chen]{liu_closing_2021}
Yong~(Alexander) Liu, Xin~(Lucy) Liu, Fang~(Nancy) Li, Haohuan Fu, Yuling Yang,
  Jiawei Song, Pengpeng Zhao, Zhen Wang, Dajia Peng, Huarong Chen, Chu Guo,
  Heliang Huang, Wenzhao Wu, and Dexun Chen.
\newblock Closing the "quantum supremacy" gap: achieving real-time simulation
  of a random quantum circuit using a new {Sunway} supercomputer.
\newblock pages 1--12, November 2021.
\newblock \href{https://doi.org/10.1145/3458817.3487399}{\ttfamily\path{
  doi:10.1145/3458817.3487399}}.
\newblock URL \url{https://doi.org/10.1145/3458817.3487399}.

\bibitem[Markidis et~al.(2018)Markidis, Der~Chien, Laure, Peng, and
  Vetter]{markidis_nvidia_2018}
Stefano Markidis, Steven~Wei Der~Chien, Erwin Laure, Ivy~Bo Peng, and
  Jeffrey~S. Vetter.
\newblock {NVIDIA} {Tensor} {Core} {Programmability}, {Performance} \&
  {Precision}.
\newblock \emph{2018 IEEE International Parallel and Distributed Processing
  Symposium Workshops (IPDPSW)}, pages 522--531, May 2018.
\newblock \href{https://doi.org/10.1109/IPDPSW.2018.00091}{\ttfamily\path{
  doi:10.1109/IPDPSW.2018.00091}}.
\newblock URL \url{http://arxiv.org/abs/1803.04014}.
\newblock arXiv: 1803.04014.

\bibitem[Markov and Shi(2008)]{markov_simulating_2008}
Igor~L. Markov and Yaoyun Shi.
\newblock Simulating {Quantum} {Computation} by {Contracting} {Tensor}
  {Networks}.
\newblock \emph{SIAM Journal on Computing}, 38\penalty0 (3):\penalty0 963--981,
  January 2008.
\newblock ISSN 0097-5397.
\newblock \href{https://doi.org/10.1137/050644756}{\ttfamily\path{
  doi:10.1137/050644756}}.
\newblock URL \url{https://epubs.siam.org/doi/10.1137/050644756}.
\newblock Publisher: Society for Industrial and Applied Mathematics.

\bibitem[Markov et~al.(2018)Markov, Fatima, Isakov, and
  Boixo]{markov_quantum_2018}
Igor~L. Markov, Aneeqa Fatima, Sergei~V. Isakov, and Sergio Boixo.
\newblock Quantum {Supremacy} {Is} {Both} {Closer} and {Farther} than {It}
  {Appears}.
\newblock \emph{arXiv:1807.10749 [quant-ph]}, September 2018.
\newblock URL \url{http://arxiv.org/abs/1807.10749}.
\newblock arXiv: 1807.10749.

\bibitem[Nguyen et~al.(2021)Nguyen, Lyakh, Dumitrescu, Clark, Larkin, and
  McCaskey]{nguyen_tensor_2021}
Thien Nguyen, Dmitry Lyakh, Eugene Dumitrescu, David Clark, Jeff Larkin, and
  Alexander McCaskey.
\newblock Tensor {Network} {Quantum} {Virtual} {Machine} for {Simulating}
  {Quantum} {Circuits} at {Exascale}.
\newblock \emph{arXiv:2104.10523 [quant-ph]}, April 2021.
\newblock URL \url{http://arxiv.org/abs/2104.10523}.
\newblock arXiv: 2104.10523.

\bibitem[Okuta et~al.(2017)Okuta, Unno, Nishino, Hido, and
  Loomis]{okuta_cupy_2017}
Ryosuke Okuta, Yuya Unno, Daisuke Nishino, Shohei Hido, and Crissman Loomis.
\newblock {CuPy}: {A} {NumPy}-{Compatible} {Library} for {NVIDIA} {GPU}
  {Calculations}.
\newblock 2017.

\bibitem[Ootomo and Yokota(2022)]{ootomo_recovering_2022}
Hiroyuki Ootomo and Rio Yokota.
\newblock Recovering single precision accuracy from {Tensor} {Cores} while
  surpassing the {FP32} theoretical peak performance:.
\newblock \emph{The International Journal of High Performance Computing
  Applications}, June 2022.
\newblock \href{https://doi.org/10.1177/10943420221090256}{\ttfamily\path{
  doi:10.1177/10943420221090256}}.
\newblock URL
  \url{https://journals.sagepub.com/eprint/C6XUQKVBS3PU5SXTTFIJ/full}.
\newblock Publisher: SAGE PublicationsSage UK: London, England.

\bibitem[Ootomo and Yokota(2023)]{ootomo_reducing_2023}
Hiroyuki Ootomo and Rio Yokota.
\newblock Reducing shared memory footprint to leverage high throughput on
  {Tensor} {Cores} and its flexible {API} extension library.
\newblock pages 1--8, February 2023.
\newblock \href{https://doi.org/10.1145/3578178.3578238}{\ttfamily\path{
  doi:10.1145/3578178.3578238}}.
\newblock URL \url{https://doi.org/10.1145/3578178.3578238}.

\bibitem[Pan and Zhang(2022)]{pan2022Simulation}
Feng Pan and Pan Zhang.
\newblock Simulation of {{Quantum Circuits Using}} the {{Big-Batch Tensor
  Network Method}}.
\newblock \emph{Physical Review Letters}, 128\penalty0 (3):\penalty0 030501,
  January 2022.
\newblock \href{https://doi.org/10.1103/PhysRevLett.128.030501}{\ttfamily\path{
  doi:10.1103/PhysRevLett.128.030501}}.

\bibitem[Pan et~al.(2022)Pan, Chen, and Zhang]{pan2022Solving}
Feng Pan, Keyang Chen, and Pan Zhang.
\newblock Solving the sampling problem of the {{Sycamore}} quantum circuits.
\newblock \emph{Physical Review Letters}, 129\penalty0 (9):\penalty0 090502,
  August 2022.
\newblock ISSN 0031-9007, 1079-7114.
\newblock \href{https://doi.org/10.1103/PhysRevLett.129.090502}{\ttfamily\path{
  doi:10.1103/PhysRevLett.129.090502}}.
\newblock \href{http://arxiv.org/abs/2111.03011}{{\ttfamily arXiv:2111.03011
  [physics, physics:quant-ph]}}.

\bibitem[Paszke et~al.(2019)Paszke, Gross, Massa, Lerer, Bradbury, Chanan,
  Killeen, Lin, Gimelshein, Antiga, Desmaison, Kopf, Yang, DeVito, Raison,
  Tejani, Chilamkurthy, Steiner, Fang, Bai, and Chintala]{paszke_pytorch_2019}
Adam Paszke, Sam Gross, Francisco Massa, Adam Lerer, James Bradbury, Gregory
  Chanan, Trevor Killeen, Zeming Lin, Natalia Gimelshein, Luca Antiga, Alban
  Desmaison, Andreas Kopf, Edward Yang, Zachary DeVito, Martin Raison, Alykhan
  Tejani, Sasank Chilamkurthy, Benoit Steiner, Lu~Fang, Junjie Bai, and Soumith
  Chintala.
\newblock {PyTorch}: {An} {Imperative} {Style}, {High}-{Performance} {Deep}
  {Learning} {Library}.
\newblock In \emph{Advances in {Neural} {Information} {Processing} {Systems}},
  volume~32. Curran Associates, Inc., 2019.
\newblock URL
  \url{https://papers.nips.cc/paper/2019/hash/bdbca288fee7f92f2bfa9f7012727740-Abstract.html}.

\bibitem[Preskill(2012)]{preskill2012Quantum}
John Preskill.
\newblock Quantum computing and the entanglement frontier, November 2012.

\bibitem[Roberts et~al.(2019)Roberts, Milsted, Ganahl, Zalcman, Fontaine, Zou,
  Hidary, Vidal, and Leichenauer]{roberts2019TensorNetwork}
Chase Roberts, Ashley Milsted, Martin Ganahl, Adam Zalcman, Bruce Fontaine,
  Yijian Zou, Jack Hidary, Guifre Vidal, and Stefan Leichenauer.
\newblock {{TensorNetwork}}: {{A Library}} for {{Physics}} and {{Machine
  Learning}}, May 2019.

\bibitem[Schlag et~al.(March 24, 2022)Schlag, Heuer, Gottesb{\"u}ren,
  Akhremtsev, Schulz, and Sanders]{schlag2022HighQuality}
Sebastian Schlag, Tobias Heuer, Lars Gottesb{\"u}ren, Yaroslav Akhremtsev,
  Christian Schulz, and Peter Sanders.
\newblock High-{{Quality Hypergraph Partitioning}}.
\newblock \emph{ACM Journal of Experimental Algorithmics}, March 24, 2022.
\newblock ISSN 1084-6654.
\newblock \href{https://doi.org/10.1145/3529090}{\ttfamily\path{
  doi:10.1145/3529090}}.

\bibitem[Schuch et~al.(2007)Schuch, Wolf, Verstraete, and
  Cirac]{schuch2007Computational}
Norbert Schuch, Michael~M. Wolf, Frank Verstraete, and J.~Ignacio Cirac.
\newblock The computational complexity of {{PEPS}}.
\newblock \emph{Physical Review Letters}, 98\penalty0 (14):\penalty0 140506,
  April 2007.
\newblock ISSN 0031-9007, 1079-7114.
\newblock \href{https://doi.org/10.1103/PhysRevLett.98.140506}{\ttfamily\path{
  doi:10.1103/PhysRevLett.98.140506}}.
\newblock \href{http://arxiv.org/abs/quant-ph/0611050}{{\ttfamily
  arXiv:quant-ph/0611050}}.

\bibitem[Suzuki et~al.(2021)Suzuki, Kawase, Masumura, Hiraga, Nakadai, Chen,
  Nakanishi, Mitarai, Imai, Tamiya, Yamamoto, Yan, Kawakubo, Nakagawa, Ibe,
  Zhang, Yamashita, Yoshimura, Hayashi, and Fujii]{suzuki_qulacs_2021}
Yasunari Suzuki, Yoshiaki Kawase, Yuya Masumura, Yuria Hiraga, Masahiro
  Nakadai, Jiabao Chen, Ken~M. Nakanishi, Kosuke Mitarai, Ryosuke Imai, Shiro
  Tamiya, Takahiro Yamamoto, Tennin Yan, Toru Kawakubo, Yuya~O. Nakagawa, Yohei
  Ibe, Youyuan Zhang, Hirotsugu Yamashita, Hikaru Yoshimura, Akihiro Hayashi,
  and Keisuke Fujii.
\newblock Qulacs: a fast and versatile quantum circuit simulator for research
  purpose.
\newblock \emph{Quantum}, 5:\penalty0 559, October 2021.
\newblock \href{https://doi.org/10.22331/q-2021-10-06-559}{\ttfamily\path{
  doi:10.22331/q-2021-10-06-559}}.
\newblock URL \url{https://quantum-journal.org/papers/q-2021-10-06-559/}.
\newblock Publisher: Verein zur Förderung des Open Access Publizierens in den
  Quantenwissenschaften.

\bibitem[Treinish et~al.(2022)Treinish, Gambetta,
  et~al.]{treinish_qiskitqiskit_2022-1}
Matthew Treinish, Jay Gambetta, et~al.
\newblock Qiskit/qiskit: {Qiskit} 0.38.0, September 2022.
\newblock URL \url{https://zenodo.org/record/2573505}.

\bibitem[Villalonga et~al.(2020)Villalonga, Lyakh, Boixo, Neven, Humble,
  Biswas, Rieffel, Ho, and Mandrà]{villalonga_establishing_2020}
Benjamin Villalonga, Dmitry Lyakh, Sergio Boixo, Hartmut Neven, Travis~S.
  Humble, Rupak Biswas, Eleanor~G. Rieffel, Alan Ho, and Salvatore Mandrà.
\newblock Establishing the quantum supremacy frontier with a 281 {Pflop}/s
  simulation.
\newblock \emph{Quantum Science and Technology}, 5\penalty0 (3):\penalty0
  034003, April 2020.
\newblock ISSN 2058-9565.
\newblock \href{https://doi.org/10.1088/2058-9565/ab7eeb}{\ttfamily\path{
  doi:10.1088/2058-9565/ab7eeb}}.
\newblock URL \url{https://doi.org/10.1088/2058-9565/ab7eeb}.
\newblock Publisher: IOP Publishing.

\end{thebibliography}

\newpage
\appendix
\section{Appendix}
\subsection{Custom kernel function for irregular shaped GEMM}
\begin{figure}[h]
    \centering
    \includegraphics[width=0.8\textwidth]{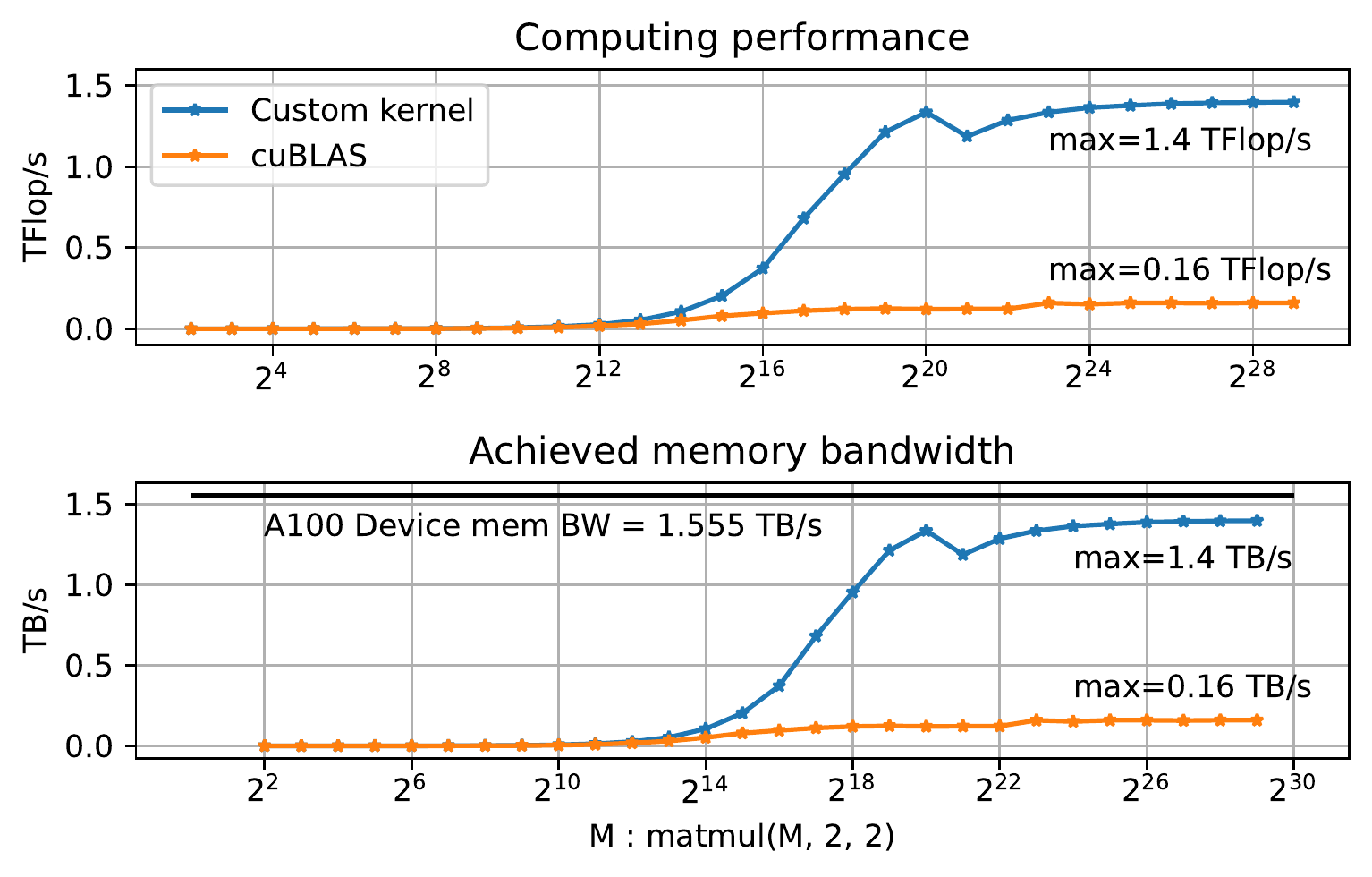}
    \caption{Throughput comparison between the custom GEMM kernel for the irregular shape of GEMMs and cuBLAS.}
    \label{fig:gemm-2xMxM}
\end{figure}
As we mention in Sec. \ref{sec:irregular-shaped-gemm}, we have implemented a custom kernel function for the irregular shape of GEMMs since the cuBLAS implementation is not optimized.
We show the throughput comparison of our GEMM and cuBLAS in Fig. \ref{fig:gemm-2xMxM}.

\end{document}